\documentclass[preprint,showpacs,preprintnumbers,amsmath,amssymb,amssymb]{revtex4}

\usepackage{graphicx}
\usepackage{dcolumn}
\usepackage{bm}
\usepackage{epstopdf}
\begin{document}
\title{Doubly charged Higgs from $e$-$\gamma$ scattering in the 3-3-1 Model}
\author{J. E. Cieza Montalvo$^1$, G. H. Ram\'{i}rez Ulloa$^2$ and M. D. Tonasse$^3$ \footnote{Permanent address: Universidade Estadual Paulista, {\it Campus} Experimental de Registro, Rua Nelson Brihi Badur 430, 11900-000 Registro, SP, Brazil}}
\affiliation{$^1$Instituto de F\'{\i}sica, Universidade do Estado do Rio de Janeiro, Rua S\~ao Francisco Xavier 524, 20559-900 Rio de Janeiro, RJ, Brazil}
\affiliation{$^2$Universidad Nacional de Trujillo, Departamento de F\'{\i}sica, Av. Juan Pablo II S/N, Ciudad Universitaria, Trujillo - Per\'u}
\affiliation{$^3$Instituto de F\'\i sica Te\'orica, Universidade Estadual Paulista, Rua Dr. Bento Teobaldo Ferraz 271, 01140-070 S\~ao Paulo, SP, Brazil}
\date{\today}



\begin{abstract}

We studied the production and signatures of doubly charged Higgs bosons in the process $\gamma e^- \rightarrow H^{--}E^+$, where $E^+$ is a heavy lepton, at the $e^-e^+$ International Linear Collider (ILC) and CERN Linear Collider (CLIC). The intermediate photons are given by the Weizs$\ddot{a}$cker-Williams and laser backscattering distributions. We found that significant signatures are obtained by bremsstrahlung and backward Comptom scattering of laser. A clear signal can be obtained for doubly charged Higgs bosons, doubly charged gauge bosons and heavy leptons.
\end{abstract}

\maketitle

\section{INTRODUCTION \label{introd}}

Nowadays it is well established that the Standard Model (SM) of electroweak interactions describes very well almost every subatomic phenomenon up to order of $\approx$ 100 GeV \cite{Aea08}. However, there is a lot of experimental results that are not well understood or nor embedded in their predictions. Among the many attempts to generalize the SM, several of them involve extensions of the Higgs sector. After spontaneous symmetry breaking, those extensions that add scalar triplet to the Higgs sector and left-right symmetry models end up with physical doubly charged Higgs bosons (DCHBs) in its particle spectrum \cite{GR81}. The DCHBs ($H^{\pm \pm}$) have some interesting consequences. The main one is related to the seesaw mechanism, which is capable of generating a small mass for the neutrino. These models are manifested in the higher energy range and the scheme is constructed so that the neutrino mass is inversely proportional to this new energy scale. This relationship can provide an idea of why the neutrino mass is so small compared with Fermi scale \cite{MS80}. One of the major motivations for the study of these particles with double charge comes from the fact that experimental physicists expect to find this kind of particles at the Large Hadron Collider (LHC), ILC and CLIC. \par

The Fermilab Tevatron can detect DCHBs via pair production if its mass does not exceed $\approx 275$ GeV an at the LHC this limit rises up to $\approx 850$ GeV. At the Tevatron, an attempt to directly detect DCHBs were carried out based on the channel $\overline{q}q \to H^{--}H^{++}$ looking for decay mode $H^{\pm\pm} \to \ell_i^\pm\ell_j^\pm$, where $\ell_{i, j}$ are known charged leptons. Four-lepton final states was observed by D$0$ Collaboration \cite{D004}, while in CDF was observed  $e^+e^+e^-e^-$, $e^+\mu^+e^-\mu^-$ and states involving $\tau$ leptons \cite{CD04}. These searches were done for left-right Model. Then, assuming a BR$\left(H^{\pm\pm} \to \mu^{\pm}\mu^\pm\right)  = 50\%$, left (right) chiral DCHBs with masses larger than 150 (136) GeV and 127 (113) GeV, respectively, are excluded \cite{Aea08}. \par

In this paper we work under the 3-3-1 electroweak model \cite{PP92,FR92,PT93}. In this class of models the electroweak symmetry gauge group SU(2)$_L$$\otimes$U(1) of the SM is extended to SU(3)$_L$$\otimes$U(1)$_N $. The simplest version of scalar sector of the model has only three Higgs triplets \cite{PT93}. Consequently, this model predicts the existence of eight
physical Higgs bosons which are four neutral bosons $h^{0}$, $H_{1}^{0}$, $H_{2}^{0}$ and $H_{3}^{0}$, a pair of single charged bosons $H^{\pm}$, and a pair of DCHBs. It loses the perturbative character on a scale of a few TeV, so that, we can admit that this is the upper limit of energy  for the model under consideration \cite{DI05}. This class of models is one of the most interesting alternatives to the SM. The process of cancellation of anomalies requires that the number of families be an integer multiple of the number of colors. But, since QCD requires that the number of colors is less than five, then in the 3-3-1 model the number of families is exactly three \cite{FR92}. \par

We are interested in production of single DCHB via interactions $\gamma e^-$ from $e^+e^-$ collisions in International Linear Collider (ILC) and in CERN Linear Collider (CLIC) \cite{chong}. The version of the 3-3-1 model which we studied predicts the existence of heavy leptons $P_{\ell a}$ $\left(P_{\ell_a} = E, M, T\right)$ \cite{PT93}. Hence, the process to be consider is $e^-\gamma \to H^{--}E^+$, which receives contributions of  $H^{--}$ and $U^{--}$ exchange in the $t$ channel and $e$ in the $s$ channel. \par

In the next Section we will present the main features of the version of the 3-3-1 Model under consideration. In Sec. III we will give the expressions for production cross sections and finally, in Sec. IV we will describe the results and conclusions.

\section{Overview of the Model \label{sec2}}

We will  summarize in this section only the most relevant points of the model.  For details see Ref. \cite{PT93}. The left-handed leptons and quarks transform under the SU(3)$_L$ gauge group as the triplets

\vskip -0.4cm

\begin{subequations}\begin{eqnarray}
\psi_{aL} & = & \left(\begin{array}{c}\nu_{\ell_a} \\ \ell^\prime_a \\ P^\prime_{\ell_a}\end{array}\right)_L \sim \left({\bf 3}, 0\right) \qquad Q_{1L} = \left(\begin{array}{c}u^\prime_1 \\ d^\prime_1 \\ J_1\end{array}\right)_L \sim \left({\bf 3}, \frac{2}{3}\right), \cr
Q_{\alpha L} & = & \left(\begin{array}{c}J^\prime_\alpha \\ u^\prime_\alpha \\ d^\prime_\alpha\end{array}\right)_L \sim \left({\bf 3^*}, -\frac{1}{3}\right),
\label{fer}\end{eqnarray}
where $\ell^\prime_a = e^\prime$, $\mu^\prime$, $\tau^\prime$ and $\alpha = 2$, $3$. The $J_1$ exotic quark carries $5/3$ units of elementary electric charge, while $J_2$ and $J_3$ carry $-4/3$ each one. In Eqs. (\ref{fer}) the numbers $0$, $2/3$, and $-1/3$ are the U(1)$_N$ charges. Each left-handed charged fermion has its right-handed counterpart transforming as a singlet in the presence of the SU(3)$_L$  group, i.e.,

\begin{eqnarray}
&& \ell_{aR}^\prime \sim \left({\bf 1}, -1\right) \qquad P_{\ell_aR}^\prime \sim \left({\bf 1}, 1\right), \qquad U^\prime_R \sim \left({\bf 1}, 2/3\right) , \cr
&& D^\prime_R \sim \left({\bf 1}, -1/3\right), \qquad J_{1R} \sim \left({\bf 1}, 5/3\right) \qquad J^\prime_\alpha \sim \left({\bf 1}, -4/3\right).
\end{eqnarray}\label{ferr}\end{subequations}
We are defining $U = u$, $c$, $t$ and $D = d$, $s$, $b$. It should be noted that, in order to avoid anomalies,  one of the quark families must transform in a different way with respect to the others. In Eqs. (\ref{ferr}) all the primed fields are linear combinations of the mass eigenstates. The charge operator is defined by

\begin{equation}
\frac{Q}{e} = \frac{1}{2}\left(\lambda_3 - \sqrt{3}\lambda_8\right) + N,
\label{op}\end{equation}
where $\lambda_3$ and $\lambda_8$ are the diagonal Gell-Mann matrices. We notice, however, that since $Q_{\alpha L}$ in Eqs. (\ref{fer}) are in antitriplet representation of SU(3)$_L$ , the antitriplet representation of the Gell-Mann matrices must also be used in Eq. (\ref{op}) in order to get the correct electric charge for the quarks of the second and third generations. \par

The three Higgs scalar triplets

\begin{equation}
\eta = \left(\begin{array}{c}\eta^0 \\ \eta^-_1 \\ \eta^+_2\end{array}\right) \sim \left({\bf 3}, 0\right), \quad \rho = \left(\begin{array}{c}\rho^+ \\ \rho^0 \\ \rho^{++}\end{array}\right) \sim \left({\bf 3}, 1\right), \quad \chi = \left(\begin{array}{c}\chi^- \\ \chi^{--} \\ \chi^0\end{array}\right) \sim \left({\bf 3}, -1\right),
\label{higgs}\end{equation}
are the minimal content of Higgs sector enough to break the symmetry spontaneously and generate the masses of fermions and gauge bosons in the model. The neutral scalar fields develop the vacuum expectation values (VEVs) $\langle\eta^0\rangle = v_\eta$, $\langle\rho^0\rangle = v_\rho$ and $\langle\chi^0\rangle = v_\chi$, with $v_\eta^2 + v_\rho^2 = 246^2$ GeV$^2$. \par

The pattern of symmetry breaking is ${\rm SU(3)}_L\otimes{\rm U(1)}_N \stackrel{\langle\chi^0\rangle} \longrightarrow {\rm SU(2)}_L\otimes{\rm U(1)}_Y \stackrel{\langle\eta^0, \rho^0\rangle} \longrightarrow {\rm U(1)}_{\rm em}$
and so, we can expect
\begin{equation}
v_\chi \gg v_\eta, v_\rho.
\label {app}\end{equation}
Then, after the symmetry breaking, the masses of scalar fields become,

\begin{subequations}\begin{eqnarray}
&& { m^2_{H_1^0} \approx 4\frac{\lambda_2v_\rho^4 - 2\lambda_1v_\eta^4}{v_\eta^2 - v_\rho^2}, \qquad m_{H_2^0}^2 \approx \frac{v_W^2}{2v_\eta v_\rho}v_\chi^2, \qquad m_{H_3^0}^2 \approx -\lambda_3v_\chi^2,}   \label{mapp}\\
&& {m_h^2 = -\frac{fv_\chi}{v_\eta v_\rho}\left[v_W^2 + \left(\frac{v_\eta v_\rho}{v_\chi}\right)^2\right], \qquad m_{H_1^\pm}^2 = \frac{v_W^2}{2v_\eta v_\rho}\left(fv_\chi - 2\lambda_7v_\eta v_\rho\right),} \\
&& {m_{H_2^\pm}^2 = \frac{v_\eta^2 + v_\chi^2}{2v_\eta v_\chi}\left(fv_\rho - 2\lambda_8v_\eta v_\chi\right), \qquad m_{H^{\pm\pm}}^2 = \frac{v_\rho^2 + v_\chi^2}{2v_\rho v_\chi}\left(fv_\eta - 2\lambda_9v_\rho v_\chi\right)}.
\end{eqnarray}\end{subequations}
where in Eqs. (\ref{mapp}) we have considered Eq. (\ref{app}). Due to the transformation properties of the fermion and the Higgs fields under SU(3)$_L$ [see Eqs. (\ref{ferr}) and (\ref{higgs})] the Yukawa interactions in the model are

\begin{subequations}\begin{eqnarray}
{\cal L}^Y_\ell & = & -G_{ab}\overline{\psi}_{aL}\ell^\prime_{bR}\rho - G^\prime_{ab}\overline{\psi^\prime}_{aL}P^\prime_{\ell_bR}\chi + {\rm H. c.}, \label{Ll}\\
{\cal L}^Y_q & = & \sum_a\left[\overline{Q}_{1L}\left(G_{1a}U^\prime_{aR}\eta + \tilde{G}_{1a}D^\prime_{aR}\rho\right) + \sum_\alpha\overline{Q}_{\alpha L}\left(F_{\alpha a}U^\prime_{aR}\rho^* + \tilde{F}_{\alpha a}D^\prime_{aR}\eta^*\right)\right] + \cr
&& + \sum_{\alpha, \beta}F^J_{\alpha\beta}\overline{Q}_{\alpha L}J^\prime_{\beta R}\chi^* + G^J\overline{Q}_{1L}J_{1R}\chi + {\rm H. c.}, \label{Lq}
\end{eqnarray}\label{yuk}\end{subequations}
where $G$'s, $\tilde{G}$'s, $F$'s, and $\tilde{F}$'s are Yukawa coupling constants with $a$, $b$ $=$ $1$, $2$, $3$ and $\alpha$, $\beta$ $=$ $2$, $3$. The interaction eigenstates, which appear in Eqs. (\ref{yuk}), can be transformed into the corresponding physical eigenstates by appropriated rotations, {\it i. e.},

\begin{equation}
\ell^\prime_{aL} = U^L_{ab}\ell_{bL}, \quad \ell^\prime_{aR} = U^R_{ab}\ell_{bR}, \quad P^\prime_{\ell_aL} = V^L_{ab}P_{\ell_bL}, \quad
P^\prime_{\ell_aR} = V^L_{ab}P_{\ell_bR}, \quad \nu_{\ell_aL} = U^{L\dagger}_{ab}\nu^\prime_{\ell_bL}.
\end{equation}

However, since the cross-section calculations imply summation on flavors (Sec. III) and the rotation matrix must be unitary, the mixing parameters have no essential effects for our purpose here. So, hereafter we suppress the prime notation for the interaction eigenstates. \par

The Lagrangian (\ref{Ll}) shows that the ordinary particle masses are proportional to $v_\rho$ and $v_\eta$, while the heavy ones are proportional to $v_\chi$. We are not considering here neutrino masses since they are not important in the discussion. \par

The gauge bosons consist of an octet $W^i_\mu$ $\left(i = 1, \dots, 8\right)$, associated with SU(3)$_L$ and a singlet $B_\mu$ associated with U(1)$_N$. The covariant derivative is

\begin{equation}
{\cal D}_\mu\varphi_a = \partial_\mu\varphi_a + i\frac{g}{2}\left(\bm{W}_\mu.\bm{\lambda}\right)^b_a\varphi_b + ig^\prime N_\varphi\varphi_aB_\mu,
\end{equation}
where $\varphi_a$ $=$ $\eta$, $\rho$, $\chi$. The model predicts single-charged $\left(V^\pm\right)$, doubly-charged $\left(U^{\pm\pm}\right)$ vector bileptons and a new neutral gauge boson $\left(Z^\prime\right)$ in addition to the charged standard gauge bosons $W^\pm$ and the neutral standard $Z$. The masses of the new gauge bosons are

\begin{equation}
m_{Z^\prime}^2 \approx \left(\frac{ev_\chi}{s_W}\right)^2\frac{2\left(1 - s_W^2\right)}{3\left(1 - 4s_W^2\right)} \quad m_V^2 = \left(\frac{e}{s_W}\right)^2\frac{v_\eta^2 + v_\chi^2}{2}, \quad m_U^2 = \left(\frac{e}{s_W}\right)^2\frac{v_\rho^2 + v_\chi^2}{2}.
\end{equation}
where $s_W = \sin\theta_W$. The charged currents are read off from

\begin{equation}
{\cal L}_C = -\frac{g}{\sqrt{2}}\sum_a\left(\overline{\ell}_{aL}\gamma^\mu\nu_{\ell_aL}W^-_\mu + \overline{P}_{\ell_aL}\gamma^\mu K_{ab}\nu_{\ell_bL}V^+_\mu + \overline{\ell}_{aL}\gamma^\mu K^\dagger_{ab}P_{\ell_bL}U^{--}\right) + {\rm c. H.},
\end{equation}
where $K = V^{L\dagger}U^L$ is a mixing matrix. \par

The main motivation of this work is to show that in the context of the 3-3-1 model of Ref. \cite{PT93}, the signatures for DCHBs can be significant at the ILC and at the CLIC. One way to search for $H^{\pm \pm}$ and $E^{\pm}$ is through the process ${e^-\gamma} \rightarrow H^{--}E^+$, where the photons comes either from bremsstrahlung or from laser backscattering. Our results indicate a satisfactory number of events to establish the signal. Analyzing it we can make inferences about the existence of DCHBs, doubly charged gauge bosons and heavy leptons.


\section{CROSS SECTION PRODUCTION}

We study the direct DCHBs production in the process $\gamma e^- \rightarrow H^{--}E^+$, which occurs through the contribution of the $s$ channel, as shown in Fig. $1$, and the contributions of the $H^{\pm\pm}$ and $U^{\pm\pm}$ in the $t$ channel (see Fig. $2$). We assume the intermediate photon is produced either by the Weizs$\ddot{a}$cker-Williams \cite{bs} or by backscattering of laser from the $e$ beam \cite{lb}. \par

The interactions Lagrangian is given Sec. \ref{sec2} and in several papers (see, for example, Ref. \cite{cnt1}). Then we evaluate the differential subprocess cross section for this reaction as

\begin{eqnarray}
\frac{d\hat{\sigma}}{d\cos\theta} & = & \frac{\beta\alpha}{8\hat{s}}\left[m_E^2(-m_E^2 + \hat{t}^2)\frac{\Lambda^2_\gamma\Lambda^2_{eE}}{\left(\hat{t} - m_{H^{\pm\pm}}^2\right)^2} + \right. \cr
&& \left. + 16\alpha^2\pi^2\frac{\Lambda^2_{U\gamma}\Lambda^2_{eEU}}{\left(\hat{t} - m_U^2\right)^2}\left(\frac{m_E^4}{m_U^2} - \frac{m_E^4}{2m_U^4}\hat{t} - \frac{m_E^2}{m_U^2}\hat{t} + \frac{m_E^2}{2m_U^4}\hat{t}^2 + m_E^2 - \hat{t}\right) +  \right. \cr
&& \left. + \frac{\Lambda^2_{eE}}{\hat{s}^2}\left(m_E^2m_{H^{\pm\pm}}^2 - m_{H^{\pm\pm}}^2\hat{u} + m_E^4 - m_E^2\hat{t} - 2m_E^2\hat{u} + \hat{u}\hat{t} + \hat{u}^2\right)\right].
\end{eqnarray}

It should be mention that the interference terms give not any contributions to this cross section and the $\Lambda_i$ are given in the form

\begin{subequations}\begin{align}
\Lambda_\gamma & = \frac{v_\chi^2 - v_\eta^2}{v_\chi^2 + v_\eta^2},  \qquad
&\Lambda_{eP} & =  \frac{i \ v_{\eta}}{2 \ v_{\chi}
\sqrt{v_{\eta}^{2}+ v_{\chi}^{2}}} + \frac{i \ v_{\chi}}{2 \ v_{\eta}
\sqrt{v_{\eta}^{2}+ v_{\chi}^{2}}}  ,   \qquad  \\
\Lambda_{U\gamma} & =  i\frac{v_\eta v_\chi}{\sin\theta_W}\sqrt{\frac{2}{v_\eta^2 + v_\chi^2}}, \qquad
& \Lambda_{UeP} & = -\frac{i}{2\sqrt{2}\sin\theta_W}.
\end{align}\end{subequations}
$\alpha$ is the fine structure constant which we take equal to  $\alpha =1/128$, $\sqrt{\hat{s}}$ is the center of mass (cm) energy of the $\gamma e^-$ system and the kinematic invariants are $\hat{t}\left(\hat{u}\right) = C_-\left(C_+\right)$ with

\begin{eqnarray}
C_{\mp} & = & m_{H^{\pm\pm}}^2 - \frac{\hat{s}}{2}\left\{1 + \frac{m_{H^{\pm\pm}}^2 - m_E^2}{s} \mp \right. \cr
&& \left. \mp \cos\theta\left[\left(1 - \frac{m_{H^{\pm\pm}}^2 + m_E^2}{s}\right)\left(1 - \frac{m_{H^{\pm\pm}}^2 - m_E^2}{s}\right)\right]^{1/2}\right\},
\end{eqnarray}
where $\theta$ is the angle between the DCHB and the incident electron in the cm-frame, in this frame $\Lambda_{\gamma}^{\mu}$ is the vertex strength of the $H^{\pm \pm}$ to $\gamma$ and $H^{\pm \pm}$ ($\Lambda_{\gamma}^{\mu} = \Lambda_{\gamma} (p_{1}- q_{1})^{\mu}$) with $q_{1}$ and $p_{1}$ being the momentum four-vectors of the $\gamma$ and $H^{\pm \pm}$, respectively.
$\Lambda_{eE}$ is the vertex strength of the $H^{\pm \pm}$ to $e^{\pm}$ and $E^{\pm}$, $\Lambda_{U\gamma}^{\mu \nu}$ is for $U^{\pm \pm}$ to  $\gamma$ and $H^{\pm \pm}$ ($\Lambda_{U\gamma}^{\mu \nu}= \Lambda_{U\gamma} g^{\mu \nu}$), $\Lambda_{UeE}$ for $U^{\pm \pm}$ to $e^{\pm}$ and $E^{\pm}$ and the others couplings are given in \cite{cnt1}. It is to note that in previous papers \cite{cnt1, cnt2, cineto, cito}, was made an error in the calculation of the coupling $H^{\pm\pm} \rightarrow e^\pm P_a^\pm$, where the index $a$ denotes the flavor, this error was now corrected ($12a$). We are assuming $\sin^2{\theta_W} = 0.2319$ \cite{Aea08}. So we obtain the total cross section for this process folding $\hat{\sigma}$ with the one photon luminosities

\begin{eqnarray}
\sigma = \int_{x_{\rm min}}^1\frac{dL}{d\tau}d\tau\hat{\sigma}\left(\hat{s} = xs\right) = \int_{x_{\rm min}}^1dxf_{\gamma/\ell}\left(x\right)\int\frac{d\hat{\sigma}}{d\cos\theta}d\cos\theta,
\end{eqnarray}
where $x_{\rm min} = \left(m_{H^{\pm\pm}} + m_E\right)^2$.

\section{RESULTS AND CONCLUSIONS}

Here we present the cross section for the process $e^{\mp} \gamma \rightarrow H^{\mp\mp}E^\pm$ for the ILC ($1.0 \ (1.5))$ TeV and CLIC ($3$ TeV). All calculations were done according to \cite{ton1,cnt2} and we obtain for the parameters and the VEV, the following representative values: $\lambda_{1} =-0.36$,  $\lambda_{2}=\lambda_{3}=-\lambda_{6}=-1$, $\lambda_{4}= 2.98$ $\lambda_{5}=-1.57$, $\lambda_{7} =-2$, $\lambda_{8}=-0.42$,  $v_{\eta}=195$ GeV, these parameters and VEV are used to estimate the values for the particle masses which are given in Table I.

\begin{table}[h]
\caption{\label{tab1} Values for the particle masses used in this work. All the values in this Table are given in GeV.}
\begin{ruledtabular}
\begin{tabular}{c|ccccccccccccccc}
$f$ & $v_{\chi}$ & $m_E$ & $m_M$ & $m_{T,{H_3^0}}$ & $m_{H^{\pm \pm}}$  & $m_{h}$ & $m_{H^\pm_1}$ & $m_{H^\pm_2}$ & $m_V$ & $m_U$ & $m_{Z^\prime}$ & $m_{J_1}$ & $m_{J_2,{J_3}}$  \\
\hline
-1008.3 & 1000 & 148.9 & 875 & 2000 & 500  &  1454.6 & 0 & 183 & 467.5 & 464  &  1707.6  & 1000  & 1410   \\
-1499.7 & 1500   & 223.3  & 1312.5  & 3000  & 500  & 2164.3 & 0 & 285.2 &  694.1 & 691.7 & 2561.3 & 1500 & 2115   \\
\end{tabular}
\end{ruledtabular}
\end{table}

It is remarkable that the cross sections were calculated for every mass $m_{H^{\pm\pm}}$ and for every $\lambda_{9}$ in such a way as to guarantee the approximation $-f \simeq v_\chi$ \cite{ton1, cnt2} (Tabel II and III).  It must be taking into consideration that the branching ratios of $H^{\pm \pm}$ are dependent on the parameters of the 3-3-1, which determines the size of various decay modes. \par

\begin{table}[h]
\caption{\label{tab2} Values for the particle masses of $H^{\pm \pm}$ which are in accord with the relation $-f \simeq v_\chi$ and the parameter $\lambda_{9}$, for $v_{\chi} = 1000$ GeV. The values of $m_{H^{\pm \pm}}$ and $f$ are given in GeV.}
\begin{ruledtabular}
\begin{tabular}{c|ccccccccccccccc}
$m_{H^{\pm \pm}}$ & 100 & 200 & 300 & 400 & 500 & 600 & 700 & 800 & 900 & 1000 &  1100 & 1200 & 1300 & 1400 & 1500 \\ \hline
$-f$ & 1004 & 1001 & 1003 &  1005 &1008 &  997 &1001 & 996 &996 & 1003 & 1010  & 1002  &995  &1011  & 995 \\
-$\lambda_{9}$ &  0.67 & 0.69  & 0.74 &0.81 &0.9 &1.0 &1.13 & 1.28 &1.44 & 1.63 & 1.84 & 2.06 &2.30 &2.57  & 2.84 \\
\end{tabular}
\end{ruledtabular}
\end{table}

\begin{table}[h]
\caption{\label{tab3} The same as Table II only instead of  $v_{\chi} = 1000$ GeV, we take $v_{\chi} = 1500$ GeV. }
\footnotesize{
\begin{ruledtabular}
\begin{tabular}{c|cccccccccccccccccc}
$m_{H^{\pm \pm}}$ & $100$ & $200$ & $300$ & $400$ & $500$ & $600$ & $700$ & $800$ & $900$ & $1000$ & $1100$ & $1200$ & $1300$ & $1400$ & $1500$ & $1600$ & $1700$ & $1800$ \\ \hline
$-f$ & $1502$ & $1505$ & $1506$ & $1499$ & $1499$ & $1503$ & $1510$ & $1496$ & $1508$ & $1499$ & $1494$  & $1491$  & $1491$  & $1502$  & $1499$ & $1508$ & $1496$ & $1509$  \\
$-\lambda_{9}$ &  $0.66$ & $0.67$  & $0.69$ & $0.72$ & $0.76$ & $0.81$ & $0.87$ & $0.93$ & $1.01$ & $1.09$ & $1.18$ & $1.28$ & $1.39$ & $1.51$ & $1.64$ & $1.78$ & $1.92$ & $2.08$  \\
\end{tabular}
\end{ruledtabular}
}
\end{table}
The figures show the behavior of the cross section of the process $e^+e^- \rightarrow e^- \gamma \rightarrow  H^{--}E^{+} + X$ as a function of $m_{H^{\pm\pm}}$ for bremsstrahlung and laser backscattering photons. In that case, the cross section for the process initiated by backscattered photons is approximately up to one order of magnitude larger than the one for bremsstrahlung photons due to the distribution of backscattered photons being harder than the one for bremsstrahlung.

\subsection{ILC Events}

So in Fig. 3 and 4, we show the cross section for the ILC for bremsstrahlung distribution. Considering that the expected integrated luminosity at the ILC will be of the order of $3.8  \times 10^5$ pb$^{-1}$/yr, then the statistics give a total of $ \simeq 1.8  \times 10^3(1.0  \times 10^2) (\simeq 5.7 \times 10^3\left(1.3 \times 10^3\right))$ events per year
to produce $H^{\pm\pm}E^\mp$ if we take $m_{H^{\pm\pm}} = 500\left(700\right)$ GeV, $v_\chi = 1000$ GeV and considering that the first two number of events  ($ \simeq 1.8  \times 10^3(1.0  \times 10^2)$) correspond to 1.0 TeV and the other two ($\simeq 5.7 \times 10^3\left(1.3 \times 10^3\right)$) to 1.5 TeV for the ILC respectively. Regarding the $v_\chi = 1500$ GeV for the same masses of $H^{\pm\pm}$ it  will give a total of $\simeq 1.6 \times 10^3\left(49\right) (\simeq 5.8 \times 10^3\left(1.3 \times 10^3 \right)$ events per year to produce the same particles, the difference between the numbers of events $49$ and $1.3 \times 10^3$ is due we have calculated near the threshold for the first number of events, see Fig. 3 and 4. It is important to remark that at the ILC with the c.m. energy $\sqrt{s}=0.5$ TeV and for the parameters chosen above we have the acceptable masses up to approximately $m_{H^{\pm\pm}} \simeq 301\left(227\right)$ GeV for $v_\chi = 1000(1500)$ GeV, \cite{cnt2}, therefore we have it not taken into account.  \par

In  respect to the backscattered photons (Fig. 5 and 6); the statistics are the following. Taking $m_{H^{\pm\pm}} = 500\left(700\right)$ GeV and $v_\chi = 1000$ GeV we will have a total of $\simeq 4.5 \times 10^4\left(2.9 \times 10^3\right) (\simeq 8.7 \times 10^4\left(3.2 \times 10^4\right))$ events of $H^{\pm\pm}E^\mp$ produced per year.  Regarding the VEV $v_\chi = 1500$ GeV and for the same $m_{H^{\pm\pm}} = 500\left(700\right)$ GeV it  will give a total of $\simeq 4.4 \times 10^4\left(7.8 \times 10^2	\right (\simeq 9.9 \times 10^4\left(3.4 \times 10^4\right))$ events per year to produce $H^{\pm \pm}E^\mp$. So, we have that for the ILC the number of events is sufficiently appreciable for both bremsstrahlung photons and backscattering photons, therefore our analysis will concentrate on the both  distributions.  \par

We are going to consider the following two promising types of signals. The first, where we consider the decay of $H^{\pm\pm}$ into two leptons pairs $e^\pm E^\pm$, taking into account that the branching ratios for these  particles would be $BR(H^{\pm \pm} \to e^{\pm} E^{\pm}) = 9.1 \times 10^{-1} \ \% (2.2 \times  10^{-1} \ \%)$ (Fig. $7$), we would have approximately $\simeq 16(0.22) \left (\simeq 52\left(3\right)  \right )$ events per year for bremsstrahlung photons, for $m_{H^{\pm\pm}} = 500\left(700\right)$ GeV, and $v_\chi = 1000$ GeV considering that the first two numbers of events  ($ \simeq 16  \ (0.22)$) correspond to 1.0 TeV and the other two ($\simeq 52 \ (3)$) to 1.5 TeV for the ILC respectively.  Regarding the $v_\chi = 1500$ GeV for the same masses of $H^{\pm\pm} = 500\left(700\right)$ GeV and considering BR$\left(H^{\pm\pm} \to e^\pm E^\pm\right) = 37.8 \ \% \left(3.9  \ \% \right)$ (Fig. 8) it  will give a total of $\simeq 6 \times 10^2 (2) \left (\simeq 2.2 \times 10^3 \left(51\right)  \right )$ events per year.

\begin{center}
\begin{table}[h]
\caption{\label{tab4}\footnotesize\baselineskip = 12pt
Results for the ILC considering decay modes for $H^{\pm\pm} \to e^\pm E^\pm$. WW and BL stands for bremsstrahlung photons using the Weisz$\ddot{a}$cker-Williams distribution and backscattered laser photons, respectively, where the first two numbers of Events/year correspond to 1.0 TeV and the other two to 1.5 TeV for the ILC }
\begin{tabular}{cccccccc}
\hline\hline
$v_\chi$ (GeV) & $m_{H^{\pm\pm}}$ (GeV) & BR$\left(H^{\pm\pm} \to e^\pm E^\pm\right)$ &  Events/year &  Mechanism  \\  \hline
$1000$ &  & $9.1\left(2.2\right) \times 10^{-1} \ \% $
  & $16(0.22) \ (52\left(3\right))$   \\ $1500$ & \raisebox{1.8ex}{$500\left(700\right)$} & $37.8\left(3.9\right) \ \% $ &  $600(2) \ (2200 \left(51\right))$ &   \raisebox{1.8ex}{WW} \\
$1000$ &  & $9.1\left(2.2\right) \times 10^{-1} \ \% $ &  $410(64) \ (791\left(70\right))$                                                              \\
$1500$ & \raisebox{1.8ex}{$500\left(700\right)$} & $37.8\left(3.9\right) \ \%$ &  $17000(30) \ (37000 \left(1300 \right)) $ & \raisebox{1.8ex}{BL} \\
\hline\hline
\end{tabular}
\end{table}
\end{center}

\vskip 0.3cm

\begin{center}
\begin{table}[h]
\caption{\label{tab5}\footnotesize\baselineskip = 12pt
Results for the ILC considering decay modes for $H^{\pm\pm} \to U^{\pm\pm}\gamma$ and $U^{\pm\pm} \to e^\pm E^\pm$. WW and BL stands for bremsstrahlung photons using the Weisz$\ddot{a}$cker-Williams distribution and backscattered laser photons, respectively, where as above the first two numbers of Events/year correspond to 1.0 TeV and the other two to 1.5 TeV for the ILC   . }
\begin{tabular}{cccccccc}
\hline\hline
$v_\chi$ (GeV) & $m_{H^{\pm\pm}}$ (GeV) & BR$\left(H^{\pm\pm} \to U^{\pm\pm}\gamma\right)$ & BR$\left(U^{\pm\pm} \to e^\pm E^\pm\right)$ & Events/year &  Mechanism  \\  \hline
$1000$ &    & $47.9\left(21.9\right) \ \% $ & $50(47.5) \ \% $ & $431(10) \ (1400 \left(140 \right)) $     \\
$1500$ & \raisebox{1.8ex}{$500\left(700\right)$} & $0.0\left(41.4\right) \ \% $ & $0.0(47.2) \ \% $ & $0.0(10) \ (0.0 \left(250 \right))$ &  \raisebox{1.8ex}{WW} \\
\cline{1-6}
$1000$ &    & $47.9\left(21.9\right) \ \%$ & $50(47.5) \ \% $ &   $11000(300) \ (21000 \left(3300 \right))$     \\ $1500$ & \raisebox{1.8ex}{$500\left(700\right)$} & $0.0\left(41.4\right) \ \% $ &  $0.0(47.2) \ \%$ & $0.0(152) \ (0.0 \left(6600 \right))$ &  \raisebox{1.8ex}{BL} \\
\hline\hline
\end{tabular}
\end{table}
\end{center}

\vskip -2.9cm

In  respect to the backscattered photons the statistics are the following. Taking $m_{H^{\pm\pm}} = 500\left(700\right)$ GeV and $v_\chi = 1000$ GeV and considering the same decay given above, for which branching ratios would be  $BR(H^{\pm \pm} \to e^{\pm} E^{\pm}) = 9.1 \times 10^{-1} \ \% (2.2 \times 10^{-1} \ \%)$, then we would have a total of  $\simeq 410(6)(\simeq 791(70))$ events per year. For $v_{\chi}=1500$ GeV and for $BR(H^{\pm \pm} \to e^{\pm} E^{\pm}) = 37.8 \ \% (3.9 \ \%)$, for the same
$m_{H^{\pm\pm}} = 500\left(700\right)$ GeV, we would have a total of $\simeq 1.7 \times 10^4 (30) (\simeq 3.7 \times 10^{4}(1.3 \times 10^{3}))$ events per year. \par

Considering that the second signal for $H^{\pm \pm}$ are $U^{\pm\pm}\gamma$ and taking into account that the BRs for these  particles would be BR$\left(H^{\pm\pm} \to U^{\pm\pm}\gamma\right) = 47.9 \ \% \left(21.9 \ \% \right)$ (Fig. 7) for $m_{H^{\pm\pm}} = 500\left(700\right)$ GeV, $v_\chi = 1000$ GeV and that $U^{\pm\pm}$ decay into $e^\pm E^\pm$, whose BR$\left(U^{\pm\pm} \to e^\pm E^\pm\right) = 50 \ \% \left(47.5 \ \% \right)$ see Fig. 9 and \cite{cineto}, then  we would have approximately $\simeq 431(10)  \ (\simeq 1.4 \times 10^3\left(1.4 \times 10^{2}  \right))$ events per year for bremsstrahlung photons, considering as above that the first two numbers of events  ($ \simeq 431  \ (10)$) correspond to 1.0 TeV and the other two ($ \simeq 1.4 \times 10^3\left(1.4 \times 10^{2}  \right)$) to 1.5 TeV for the ILC. Regarding $v_\chi = 1500$ GeV it will not give any event for $m_{H^{\pm\pm}} = 500$ GeV because it its restricted by the values of $m_U$ which in this case give $691.8$ GeV (Table \ref{tab1}). Taking the BR$\left(H^{\pm\pm} \to U^{\pm\pm}\gamma\right) = 41.4 \ \% $ (Fig. 7), and BR$\left(U^{\pm\pm} \to e^\pm E^\pm\right) = 47.2 \ \% $ see Fig. 10 and  \cite{cineto}, for the mass of Higgs equal to $m_{H^{\pm\pm}} = 700$ GeV and the same $v_{\chi}$ then we have for the number of events per year a total of $\simeq 10 \ (\simeq 2.5 \times 10^{2})$ given that the first one number of events  ($ \simeq 10$) correspond to 1.0 TeV and the other ($ \simeq 2.5 \times 10^{2}$) to 1.5 TeV for the ILC. Considering the backscattered photons, and taking into account that the BR$\left(H^{\pm\pm} \to U^{\pm\pm}\gamma\right) = 47.9 \ \% \left(21.9 \ \% \right)$ (Fig. 7) and BR$\left(U^{\pm\pm} \to e^\pm E^\pm\right) = 50 \ \% \left(47.5 \ \% \right)$ (Fig. 9 and \cite{cineto}), then the number of events per year will be $\simeq 1.1 \times 10^4\left(3 \times 10^2\right) (\simeq 2.1 \times 10^4\left(3.3 \times 10^3\right))$, for the masses of the Higgs boson $m_{H^{\pm\pm}} = 500\left(700\right)$ GeV and $v_\chi = 1000$ GeV.  Regarding $v_\chi = 1500$ GeV, it will  not give any event for $m_{H^{\pm\pm}} = 500$ GeV because the same reasons given above. Considering the same branching ratios as above, that is BR$\left(H^{\pm\pm} \to U^{\pm\pm}\gamma\right) = 41.4 \ \% $ (Fig. 7), and BR$\left(U^{\pm\pm} \to e^\pm E^\pm\right) = 47.2 \ \% $ (Fig. 10 and \cite{cineto}) for the mass of Higgs equal to $m_{H^{\pm\pm}} = 700$ GeV and the same $v_{\chi}$ then the number of events per year will be $\simeq 152 \ (\simeq 6.6 \times 10^{3})$. All these results are resumed in Table IV and Table V.

\subsection{CLIC Events}

Considering that the expected integrated luminosity for the CLIC will be of the order of $3 \times 10^6$ pb$^{-1}$/yr. Then the statistics we are expecting for this collider for bremsstrahlung distribution (Fig. 11) are a total of $\simeq 9.9 \times 10^4\left(\simeq 3.9 \times 10^4\right)$ of $H^{\pm\pm}$ and $E^\pm$ particles produced per year if we take the mass of the boson $m_{H^{\pm\pm}} = 500\left(700\right)$ GeV and $v_\chi = 1000$ GeV. In respect to the $v_\chi = 1500$ GeV for the same $m_{H^{\pm\pm}} = 500\left(700\right)$ GeV it  will give a total of $\simeq 1.1 \times 10^5\left(\simeq 4.2 \times 10^4\right)$ events per year to produce the same particles. \par

Taking the same types of signals as above, that is, BR$\left(H^{\pm\pm} \to e^\pm E^\pm\right) = 9.1 \times 10^{-1} \ \% \left(2.2 \times 10^{-1} \ \% \right)$ (Fig. 7), we would have approximately $\simeq 9 \times 10^2\left(\simeq  86\right)$ events per year for $m_{H^{\pm\pm}} = 500\left(700\right)$ GeV and $v_\chi = 1000$ GeV. Regarding  $v_\chi = 1500$ GeV and considering the same parameter as above and taking BR$\left(H^{\pm\pm} \to e^\pm E^\pm\right) = 37.8 \ \% \left(3.9 \ \% \right)$ (Fig. 8) then we would have $\simeq 4.2 \times 10^4\left(\simeq 1.6 \times 10^3\right)$ events per year, for $m_{H^{\pm\pm}} = 500\left(700\right)$ GeV. For the second signal, considering $m_{H^{\pm\pm}}=500\left(700\right)$ and $v_\chi = 1000$ GeV, for which BR$\left(H^{\pm\pm} \to U^{\pm\pm}\gamma\right) = 47.9 \ \% \left(21.9 \ \% \right)$ (see Fig. 7) and BR$\left(U^{\pm\pm} \to e^\pm E^\pm\right) = 50 \ \% \left(47.5 \ \% \right)$ (Fig. 9 and \cite{cineto}), it will give $ \simeq 1.8 \times 10^4\left(\simeq 4.1 \times 10^3\right) $ events per year, for bremsstrahlung photons. In respect to $v_\chi = 1500$ GeV, it will not give any event due to the same considerations given above, except in the case of $m_{H^{\pm \pm}}=700$ GeV, which branching ratios are  BR$\left(H^{\pm\pm} \to U^{\pm\pm}\gamma\right) = 41.4 \ \% $ (Fig. 7), and BR$\left(U^{\pm\pm} \to e^\pm E^\pm\right) = 47.2 \ \% $ (Fig. 10 and \cite{cineto}) for $m_{H^{\pm\pm}} = 700$ GeV and the same $v_{\chi}$, then the number of events per year will be $\simeq 8.2 \times 10^{3}$. All these results are resumed in Table VI and Table VII. \par

\begin{center}
\begin{table}[h]
\caption{\label{tab6}\footnotesize\baselineskip = 12pt
The same as in Table \ref{tab4}, but for the CLIC collider. }
\begin{tabular}{cccccccc}
\hline\hline
$v_\chi$ (GeV) & $m_{H^{\pm\pm}}$ (GeV) & BR$\left(H^{\pm\pm} \to e^\pm E^\pm\right)$ &  Events/year &  Mechanism  \\ \hline
$1000$ &   & $9.1\left(2.2\right) \times 10^{-1} \ \% $ &  $9.2 \times 10^{2}\left(86\right)$       \\
$1500$ & \raisebox{1.8ex}{$500\left(700\right)$} & $37.8\left(3.9\right) \ \% $ & $42 \left(1.6\right) \times 10^{3} $  &   \raisebox{1.8ex}{WW} \\
$1000$ &    & $9.1\left(2.2\right) \times 10^{-1} \ \%$ &  $46\left(7.2\right) \ \times 10^{2}$  \\
$1500$ & \raisebox{1.8ex}{$500\left(700\right)$} & $37.8\left(3.9\right) \ \%$ &  $22 \left(1.5 \right) \times 10^4 $ & \raisebox{1.8ex}{BL} \\
\hline\hline
\end{tabular}
\end{table}
\end{center}

\begin{center}
\begin{table}[h]
\caption{\label{tab7}\footnotesize\baselineskip = 12pt
The same as in Table \ref{tab5}, but for the CLIC collider. }
\begin{tabular}{cccccccc}
\hline\hline
$v_\chi$ (GeV) & $m_{H^{\pm\pm}}$ (GeV) &  BR$\left(H^{\pm\pm} \to U^{\pm\pm}\gamma\right)$ & BR$\left(U^{\pm\pm} \to e^\pm E^\pm\right)$ & Events/year &  Mechanism  \\ \hline
$1000$ &    & $47.9\left(21.9\right) \ \% $ & $50(47.5) \ \% $ & $18 \left(4.1 \right) \times 10^3 $      \\
$1500$ & \raisebox{1.8ex}{$500\left(700\right)$} & $0.0\left(41.4\right) \ \% $ & $0.0(47.2) \ \% $ & $0.0 \left(8.2 \times 10^ {3})\right)$ &  \raisebox{1.8ex}{WW} \\
\cline{1-6}
$1000$ &      & $47.9\left(21.9\right) \ \%$ & $50(47.5) \ \%$ &   $13 \left(3.4 \right) \times 10^4$    \\
$1500$ & \raisebox{1.8ex}{$500\left(700\right)$} & $0.0\left(41.4\right) \ \% $ &  $0.0(47.2) \ \%$ &  $0.0 \left(7.6 \times 10^{4}\right)$ &  \raisebox{1.8ex}{BL} \\
\hline\hline
\end{tabular}
\end{table}
\end{center}

Referring to the backscattered photons (Fig. 12), considering $m_{H^{\pm\pm}} = 500\left(700\right)$ GeV and $v_\chi = 1000$ GeV, we will have a total of $\simeq 5.1 \times 10^5\left(3.3 \times 10^5\right)$ events of $H^{\pm\pm}E^\mp$ produced per year.  Regarding $v_\chi = 1500$ GeV and for the same masses above  it will give a total of $\simeq 5.7 \times 10^5\left(3.9 \times 10^5\right)$ events per year to produce $H^{\pm\pm}E^\mp$. Taking the $m_{H^{\pm\pm}} = 500\left(700\right)$ GeV and $v_\chi = 1000$ GeV and taking the same first signal as above, that is, BR$\left(H^{\pm\pm} \to e^\pm E^{\pm}\right) = 9.1 \times 10^{-1} \ \% \left(2.2 \times 10^{-1} \ \% \right)$ (Fig. 7) we will have a total of $\simeq 4.6 \times 10^3\left(7.2 \times 10^2\right)$ events per year. With respect to $v_\chi = 1500$ GeV, and $m_{H^{\pm\pm}} = 500\left(700\right)$ GeV, and considering now BR$\left(H^{\pm\pm} \to e^\pm E^\pm\right) = 37.8 \ \% \left(3.9 \ \% \right)$ (Fig. 8) it will give a total of $\simeq 2.2 \times 10^5\left(1.5 \times 10^4\right)$ events per year. Considering the second signal for $m_{H^{\pm\pm}} = 500(700)$ GeV and $v_\chi = 1000$ GeV and taking into account that BR$\left(H^{\pm\pm} \to U^{\pm\pm}\gamma\right) = 49.9 \ \% \left(21.9 \ \%  \right)$ (Fig. 7), and BR$\left(U^{\pm\pm} \to e^\pm E^\pm\right) = 50 \ \%  \left(47.5 \ \%  \right)$ ( Fig. 9 and \cite{cineto}),  we would have $\simeq 1.3 \times 10^5\left(3.4 \times 10^4\right)$ events per year. Considering $v_\chi = 1500$ GeV it will not given any event for  $m_{H^{\pm\pm}} = 500$ for the same reasons given above. For $m_{H^{\pm\pm}} = 700$ it presents $7.6 \times 10^{4}$ events per year regarding the branching ratios  BR$\left(H^{\pm\pm} \to U^{\pm\pm}\gamma\right) = 41.4 \ \% $ (Fig. 7) and BR$\left(U^{\pm\pm} \to e^\pm E^\pm\right) = 47.2 \ \% $ (Fig 10 and \cite{cineto}). All these results are resumed in Table VI and Table VII. \par

In this way we have as a signal for the process $e^{\mp} \gamma \rightarrow  H^{\mp\mp} E^{\pm} \to e^\mp E^\mp E^{\pm}$, this final state can be considered as the golden mode of decay. The background wich can appear is $e^{-}  \gamma  \rightarrow e^{-}  \rightarrow Z e^{-} \rightarrow e^{-}e^{+}e^{-}$, but these backgrounds can be readely reduced by impossing the Z window cut where the invariant mass of opposite-sign lepton pairs must be far from the Z mass: \ $|m_{e^\mp E^\pm}-m_{Z}|> 10$ GeV , this removes events where the leptons come from the Z decay \cite{cheng}. \par

These two charged leptons $e^{\mp} E^{\pm}$ are relatively easy to measure in the detector, as they leave tracks and do not shower like gluons and quarks, by computing the invariant masses of both the pairs. On the other side the signal of a process $ E^{\pm} H^{\mp \mp} \to  E^{\pm} U^{\mp \mp}\gamma$ and $U^{\mp \mp} \to e^\mp E^\mp$, are $ E^{\pm} e^{\mp} E^{\mp}  \gamma$, where the $\gamma$ and $E^{\mp}$ will be measured with high accuracy by the detector and a sharp doubly charged gauge bosons will be observed in the two same-sign lepton invariant mass distribution.
If we see this signal we will not only be seeing the DCHBs but also the doubly charged gauge bosons and heavy leptons. So the $e^\pm\gamma$ collisions can be also a plentiful source of DCHBs. \par

The discovery of DCHBs will be without any doubt of great importance for the physics beyond the SM because of the confirmation of the Higgs triplet representation and the verification of the existence of some exotic particles. \par

In summary, through this work, we have shown that in the context of the 3-3-1 Model the signatures for DCHBs can be significant for $e^\pm\gamma$ collisions obtained by bremsstrahlung and backward Compton scattering of laser. This kind of new particles have distinct experimental signals of like-sign dileptons, offering excellent potential for DCHBs  discovery. Their observation in future high energy collider experiments would be a clear evidence of new physics. Thus, searching for $H^{\pm \pm}$ is one of the main goals of current and future high energy collider experiments. Our study indicates the possibility of obtaining a clear signal of these new particles
with a satisfactory number of events. \par

\acknowledgments
One of us (J. E. C. M.) would like to thank to Prof. O. J. P. \'Eboli for the proposal of this work and to hospitality of the Departamento de F\'isica Matem\^atica -USP-Brazil, where part of this work was done and M. D. T. is beholden to Instituto de F\'\i sica Te\'orica of the UNESP for his hospitality and to Conselho Nacional de Desenvolvimento Cient\'\i fico e Tecnol\'ogico for partial support. This work was supported by Funda\c c\~ao de Amparo \`a Pesquisa no Estado de S\~ao Paulo (Processo No. 2009/02272-2).


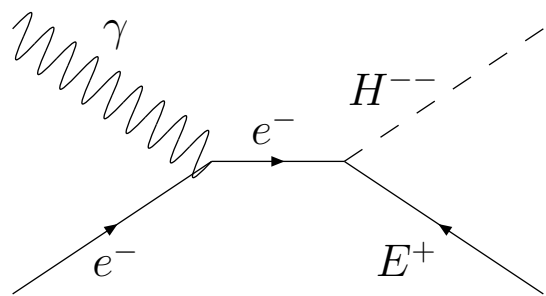
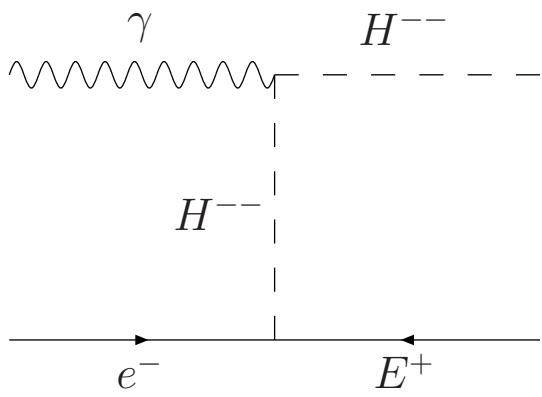
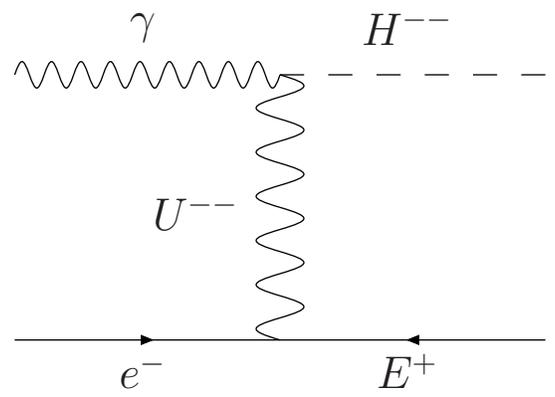
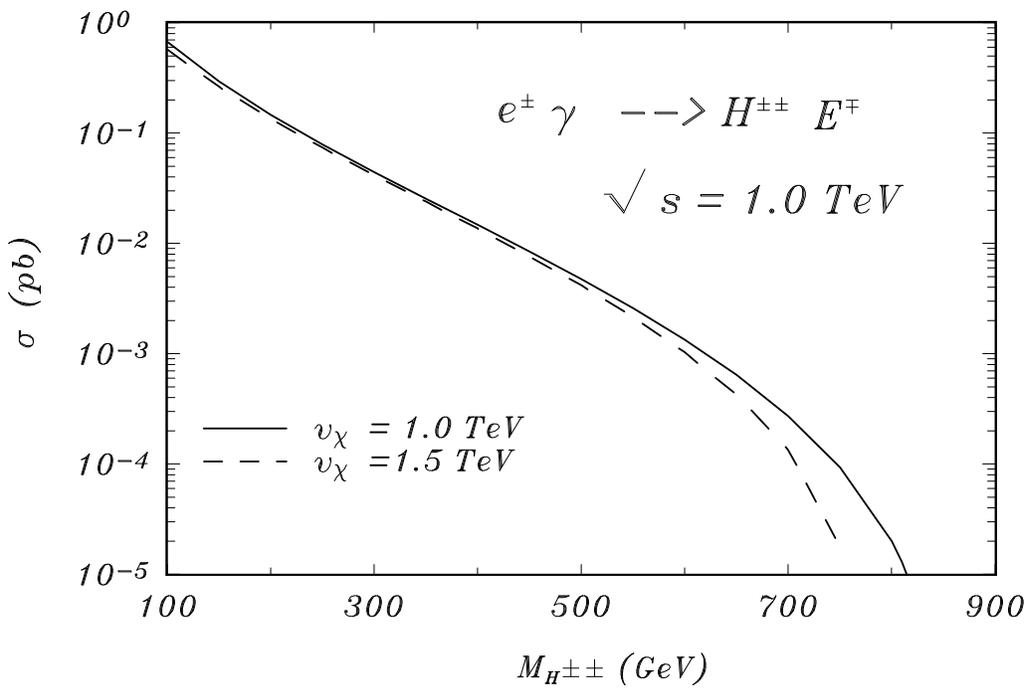
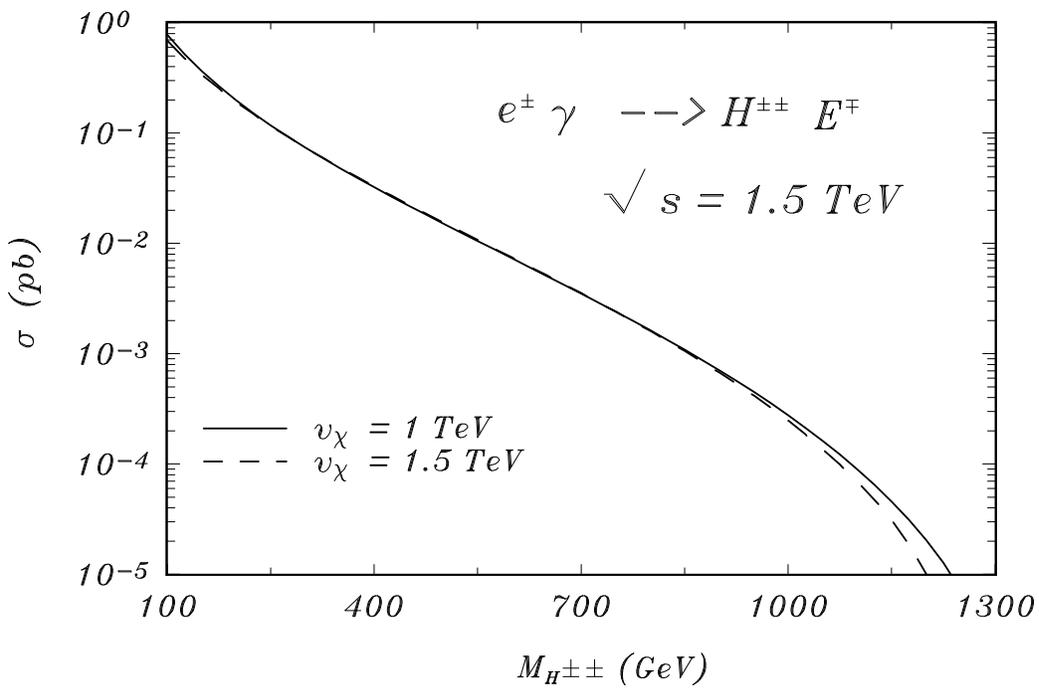
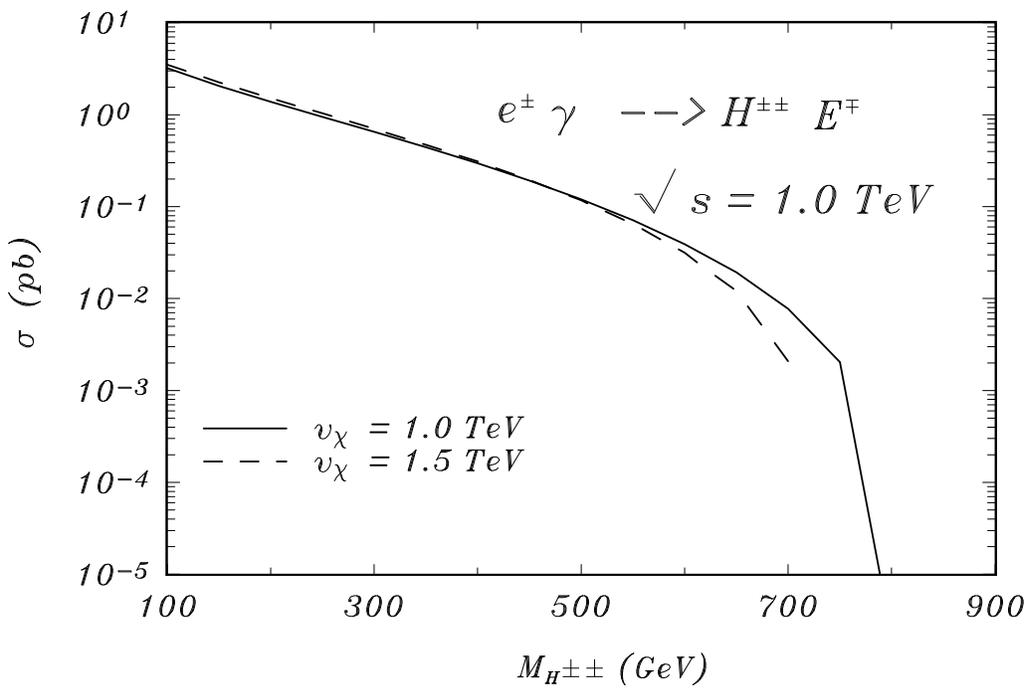
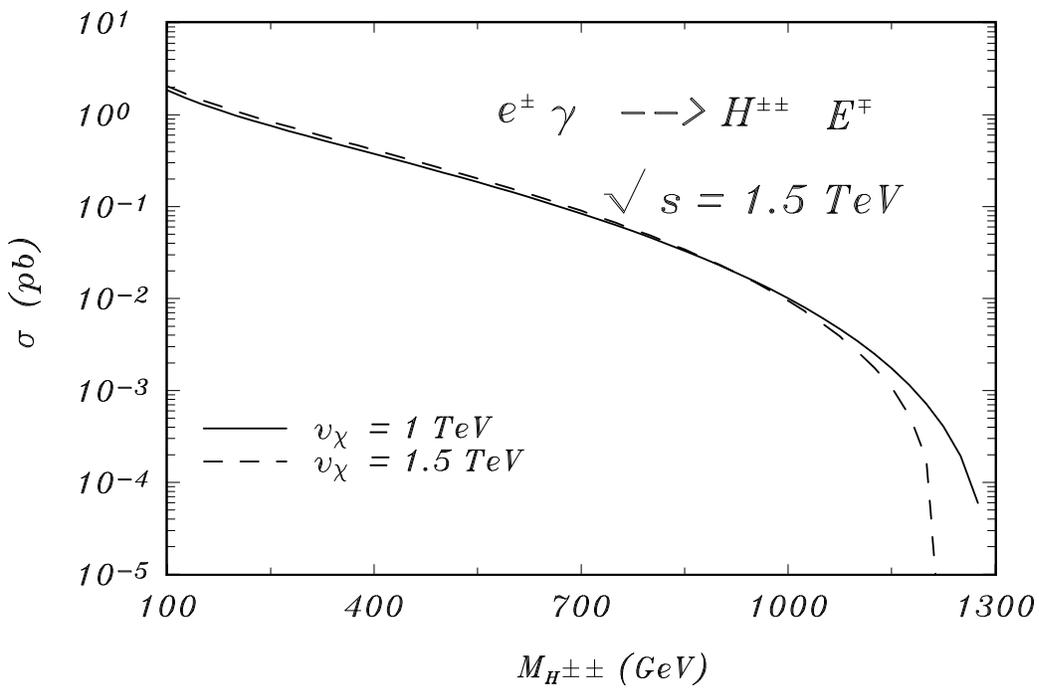
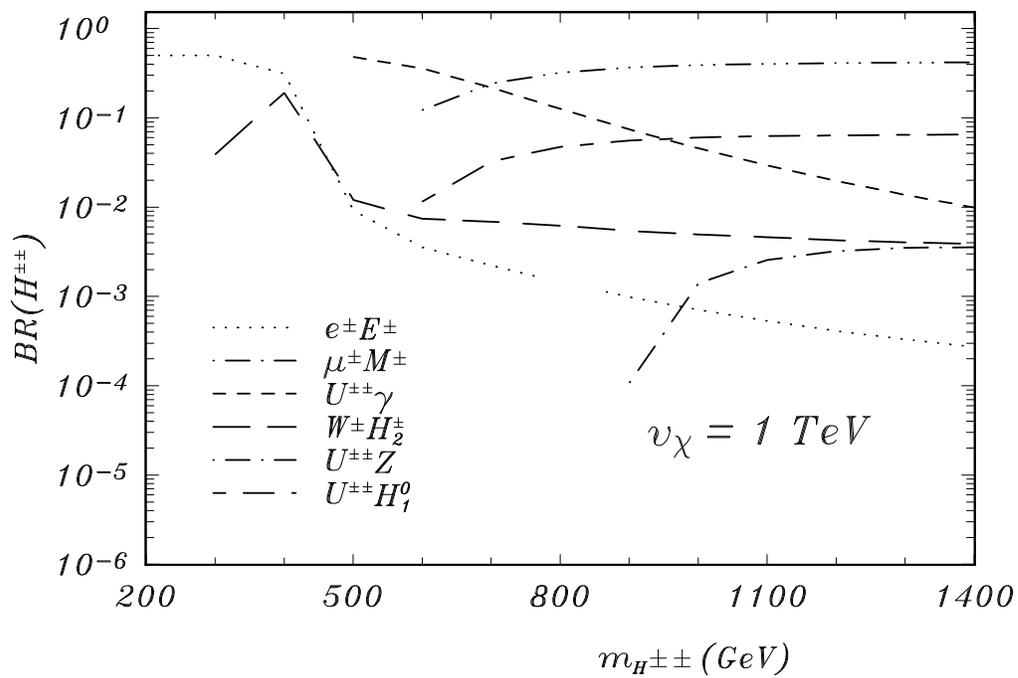
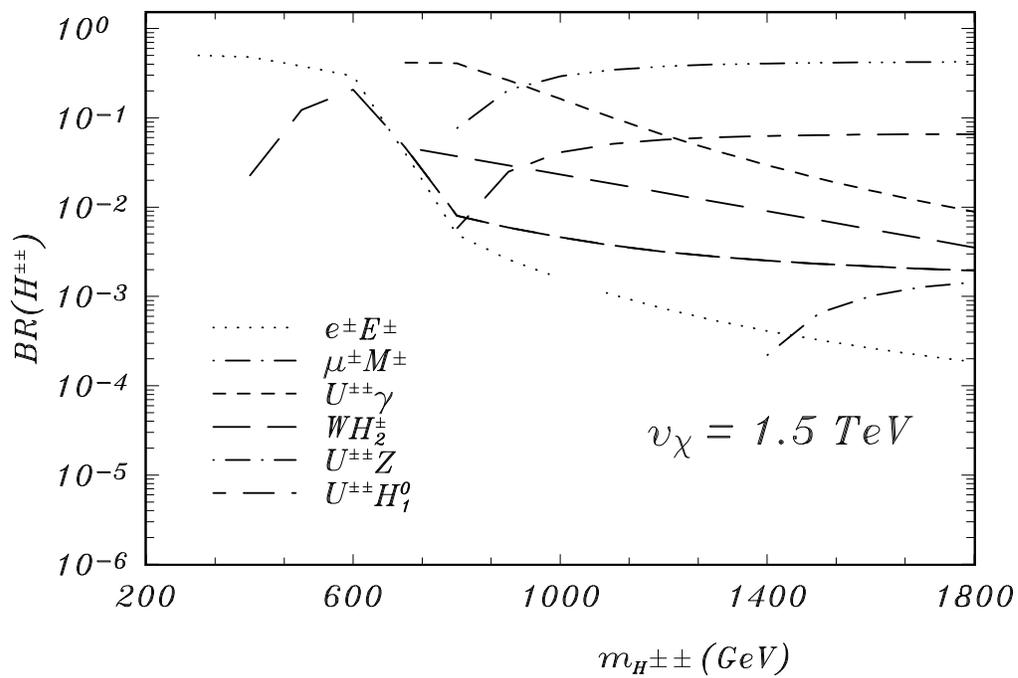
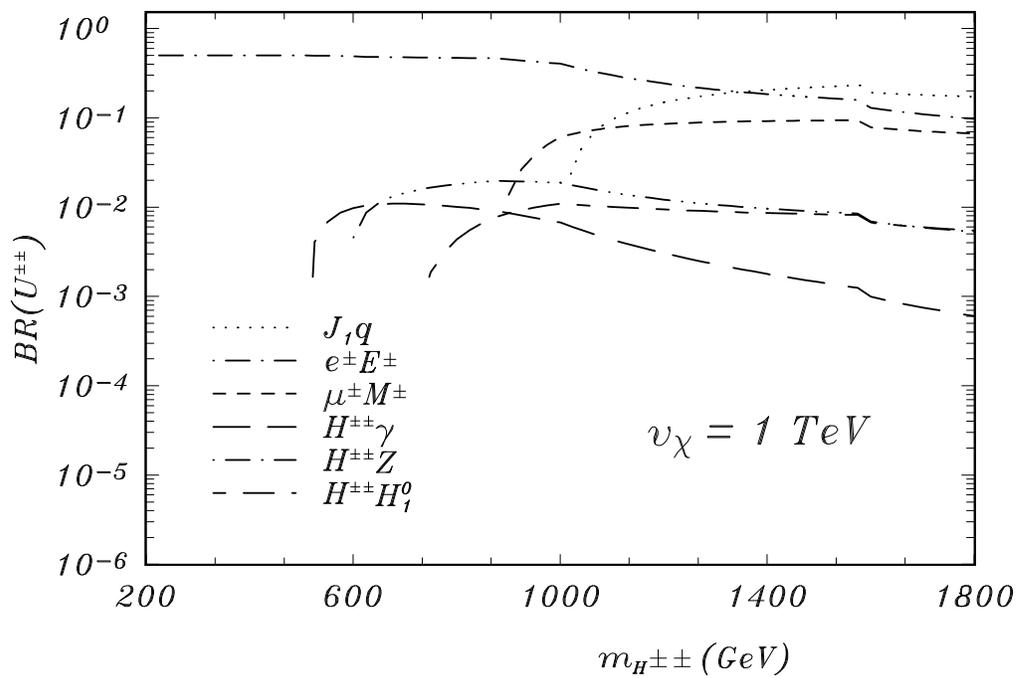
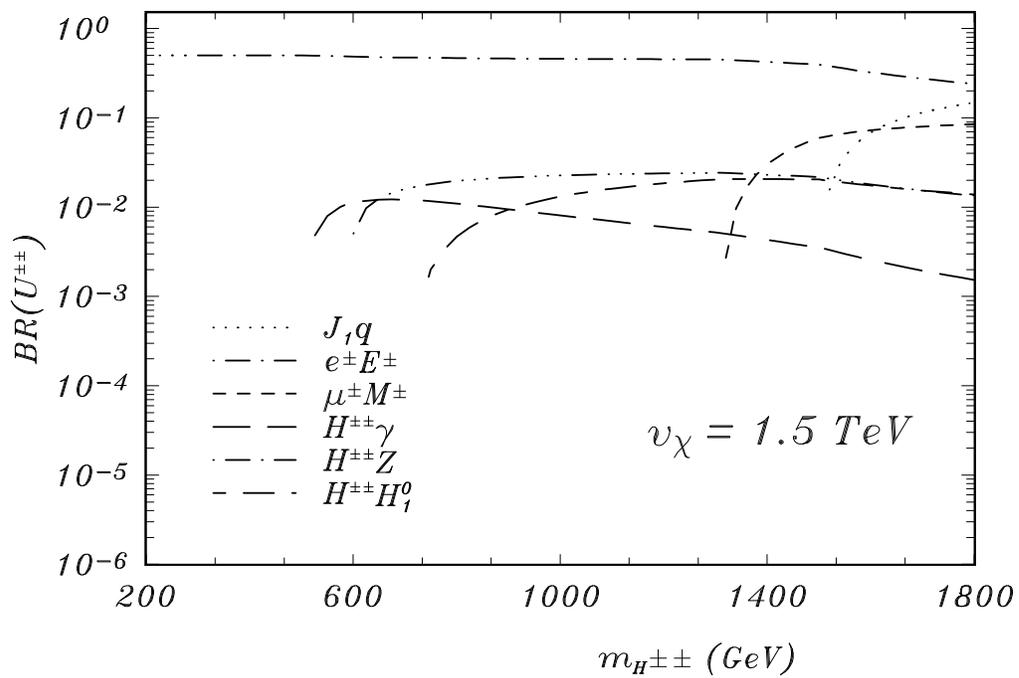
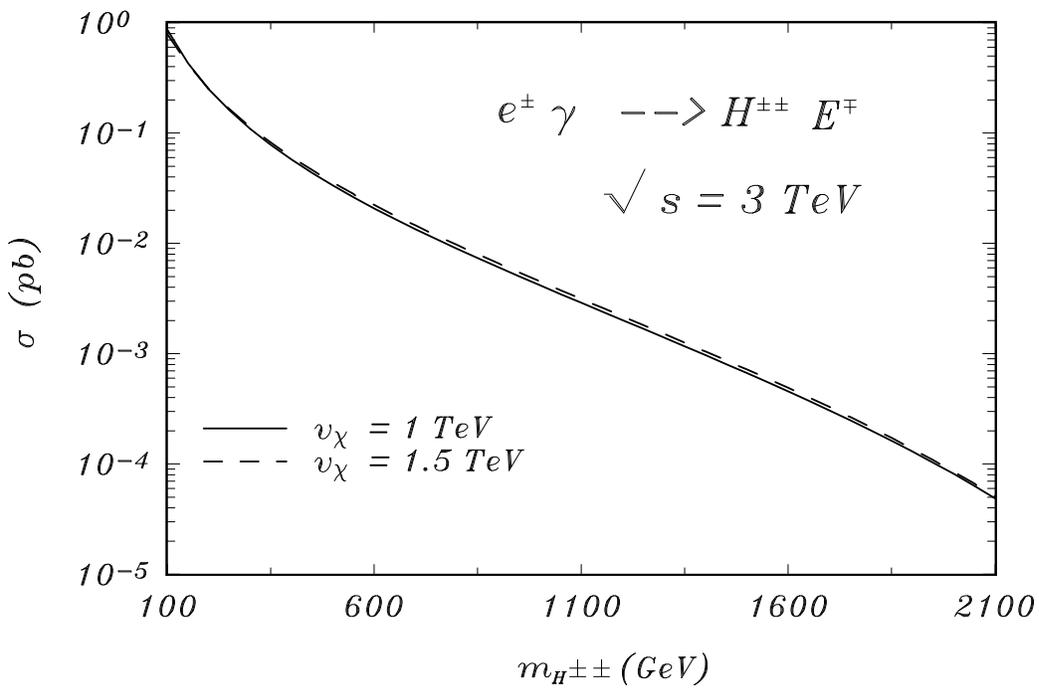
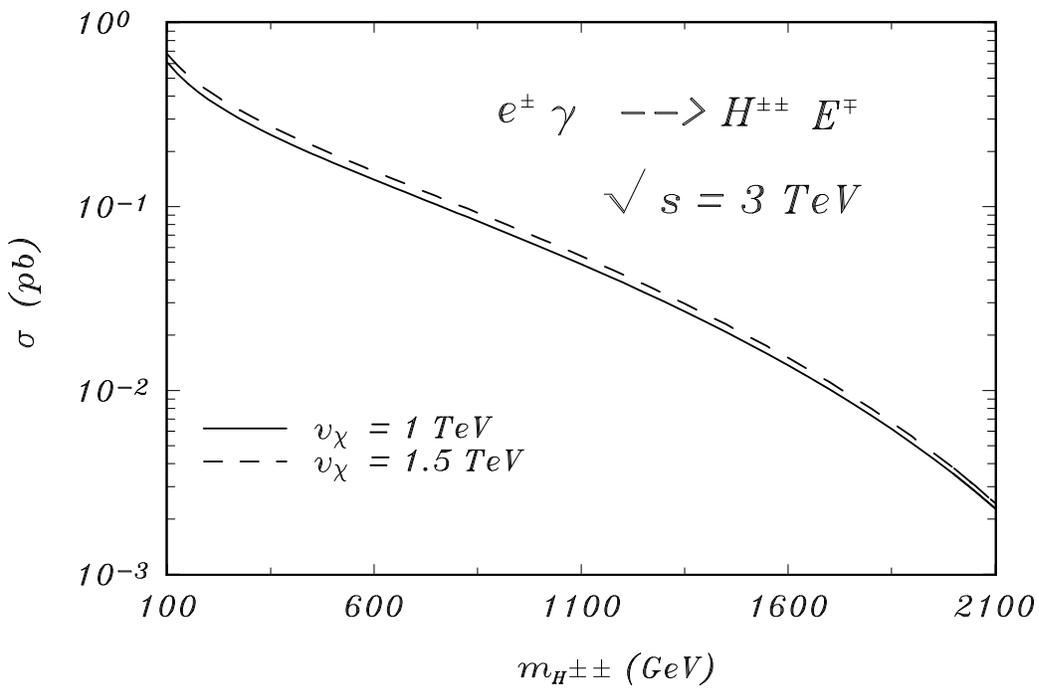
\end{document}